# Vers la conception de moyens et méthodes fondés sur les modèles pour caractériser et diagnostiquer la maturité numérique des enseignants

*Towards the design of model-based means and methods to characterize and diagnose teachers' digital maturity*

Christine MICHEL[1] ; Laëtitia PIERROT[2]

[1]Université de Poitiers, UR TECHNE, Poitiers, France
[2]Le Mans Université, UR CREN, Le Mans, France

**Résumé.** Cet article traite de la manière de combiner les modèles de maturité numérique des enseignants pour en proposer une version unifiée, utilisable pour concevoir des moyens et méthodes de diagnostic. Onze modèles applicables au champ de l'enseignement scolaire obligatoire ont été identifiés par une revue de littérature. Les modèles et la manière dont leurs dimensions constitutives contribuent à déterminer des niveaux de maturité ont été analysées. La synthèse réalisée met en lumière la diversité des dimensions utilisées, et une prise en compte partielle de la maturité numérique. De plus, la plupart de ces modèles se concentrent sur les derniers niveaux de maturité, associés à des enseignants innovateurs ou pionniers. Les modèles ont tendance à écarter les enseignants non-utilisateurs ou faibles utilisateurs numériques, pourtant représentés dans le contexte français. Dans la dernière partie de l'article, une proposition de modèle unifié de la maturité numérique des enseignants, MUME, répondant à ces deux problèmes, est décrite ainsi que les résultats préliminaires d'une étude visant à concevoir une méthode de diagnostic.
**Mots-clés :** modèle de maturité numérique, intégration du numérique à l'école, pratiques numériques des enseignants

*Abstract. This article examines how models of teacher digital maturity can be combined to produce a unified version that can be used to design diagnostic tools and methods. 11 models applicable to the field of compulsory education were identified through a literature review. The models and how their constituent dimensions contribute to the determination of maturity levels were analyzed. The summary highlights the diversity of the dimensions used and the fact that digital maturity is only partially taken into account. What's more, most of these models focus on the most recent maturity levels associated with innovative or pioneering teachers. The models tend to exclude teachers who are not digital users or who have a low level of digital use, but who are present in the French context. In the final part of the article, a proposal for a unified model of teachers' digital maturity, MUME, which addresses these two issues, is described, together with the preliminary results of a study aimed at designing a diagnostic method.*

# 1. INTRODUCTION

Dans un contexte d'émergence et de diffusion d'innovations technologiques, la transformation numérique est l'un des enjeux les plus critiques en éducation (Antonietti *et al.*, 2023). Ce processus exige des enseignants des capacités à s'adapter et être préparé à intégrer de nouveaux outils, constamment renouvelés (McCarthy *et al.*, 2023). En France, la transformation est d'autant plus cruciale que les rapports sur les usages des enseignants du premier comme du second degré témoignent d'une faible intégration des outils numériques[1]. Le potentiel d'intégration des outils numériques dans l'enseignement et l'apprentissage ne dépend pas principalement du type de technologie ou de sa fréquence d'utilisation, mais plutôt de la façon dont ces outils numériques sont utilisés pour stimuler cognitivement les élèves et les engager dans des activités d'apprentissage (Antonietti *et al.*, 2023). Et, bien que la crise sanitaire de 2020 ait eu un effet de stimulation sur les usages numériques, ils restent encore limités à des pratiques de communication et transmission de ressources (Michel et Pierrot, 2022 ; Plantard et Serreau, 2024).

Avec l'objectif de décrire le déploiement des technologies en éducation, différents modèles sont proposés. Par exemple, le TPACK (Mishra et Koehler, 2006), le SAMR (Puentedura, 2012), le NETS-T (ISTE, 2017) ou le DigCompEdu (Redecker, 2017) abordent les différentes dimensions de l'activité professionnelle d'un enseignant. Ces modèles traitent plus globalement des dynamiques d'intégration et des niveaux de maturité numérique des enseignants. Les niveaux de maturité numérique représentent un moyen de mesurer la manière dont la technologie est utilisée pour transformer et redéfinir les activités professionnelles de l'enseignant, au-delà de considérations quantitatives (fréquences d'usage par exemple). Ces niveaux tiennent en effet compte de la qualité de l'enseignement, comme le type d'activités mises en œuvre (Backfisch *et al.*, 2021).

Les modèles existants sont relativement hétérogènes dans leur manière de considérer la nature des activités professionnelles ou les niveaux de maturité. Par exemple, le modèle TPACK (Mishra et Koehler, 2006) décrit les différents champs de connaissances (pédagogique, didactique et technique) mobilisées par les enseignants, lorsqu'ils utilisent des technologies dans des activités de formation. Si ce modèle présente l'avantage majeur de proposer une conceptualisation sommaire de l'activité de formation, cette simplification ne permet pas de faire état de ses spécificités (par exemple, s'agit-il d'une activité d'apprentissage ? d'enseignement ? à quelles fins ? selon quel format ?), ni des habiletés précises attendues pour la mettre en œuvre. Cependant, les modèles identifiés sont comparables dans le sens où ils n'incluent pas de descriptions liées à la maîtrise de technologies particulières, de manière à pouvoir rester valides alors que l'innovation technologique est en constante évolution.

Dans cette perspective, un modèle de maturité numérique peut être mobilisé pour soutenir les individus dans le processus de transformation numérique (McCarthy *et al.*, 2023). En effet, un tel modèle décrit les différentes dimensions de l'activité professionnelle nécessaires ou touchées par le processus de transformation. La maturité numérique y est définie comme la capacité d'un individu à s'emparer des technologies de manière à soutenir son propre développement personnel et à s'intégrer socialement (Laaber *et al.*, 2023). Rapportée au contexte éducatif, la vision globale des modèles de maturité numérique peut guider les enseignants ou l'équipe d'encadrement dans le choix des pratiques professionnelles à développer. L'enjeu principal est alors de faire en sorte d'aligner ces pratiques relatives à l'utilisation de technologies à des objectifs de réussite scolaire (à travers l'implication des élèves ou la personnalisation des apprentissages par exemple). Dit autrement, les modèles de maturité numérique en éducation s'intéressent aux différentes dimensions qui agissent sur l'intégration des technologies, en particulier le pilotage des actions de numérisation des structures et l'activité professionnelle des enseignants.

---

[1] https://www.education.gouv.fr/media/95365/download

Au-delà des problématiques d'accès, de disponibilité et de fréquence d'usage, la maturité numérique doit prendre en considération, dans ces modèles, des questions de politique institutionnelle et de pédagogie (par exemple au niveau des contenus disciplinaires et pédagogiques traités, ou encore des changements attendus en ajoutant l'outil numérique) soulevées par l'introduction des technologies (Franklin et Bolick, 2007). Les modèles de maturité trouvent aussi leur intérêt pour mesurer, diagnostiquer ou accompagner les enseignants dans leur utilisation du numérique (Kimmons *et al.*, 2020). Plus largement, l'étude de la maturité à travers des modèles permet d'aborder l'adoption des outils numériques en combinant des facteurs liés à l'enseignant, et à son contexte d'exercice professionnel (Harrison *et al.*, 2014). Pour cela, il est nécessaire de considérer l'apprenant, l'enseignant et le contexte d'utilisation plus large en collectant des données pour mesurer l'étendue et la profondeur de l'intégration du numérique dans un établissement (Underwood *et al.*, 2010 ; Underwood et Dillon, 2004). Cette approche socio-contextuelle de l'adoption du numérique se distingue des travaux portant sur le « bon » usage attendu du numérique, et invite à s'intéresser notamment aux enseignants, en tant « qu'agents de changement », voire de « leader » qui mettent en œuvre l'outil (ISTE, 2017 ; Leite et Lagstedt, 2021). L'état de l'art réalisé dans le cadre de cet article s'inscrit dans cette perspective.

La multitude de modèles existants sur la maturité numérique des enseignants met en concurrence des observations empiriques de pratiques conceptualisées, des propositions théoriques non éprouvées sur le terrain et d'autres validés empiriquement. Or, ces modèles deviennent les socles à partir desquels se font des analyses empiriques, et des stratégies de formation des enseignants ou de diagnostics des établissements. Ces modèles sont aussi utiles pour construire des cursus de formation ou adapter les EIAH au profil des apprenants, qu'ils s'agissent d'élèves ou d'enseignants. En effet, tout comme le CRCN (adapté du DigComp) et Pix ont structuré les méthodes de certification et formation au numérique des élèves en France, les enseignants seront prochainement certifiés sur la base du CRCN-Edu (adaptation du DigCompEdu) et formés par une version spécifique de Pix : Pix+Edu[2].

L'objectif de notre étude est donc, à partir d'une revue de littérature, de faire une analyse des différents modèles de maturité et de proposer une version unifiée ayant une dimension holistique. Notre question générale de recherche (QR) est la suivante : quel modèle représente le mieux la maturité numérique des enseignants ? Spécifiquement, selon quels domaines de la définir ? (QR1) Selon quels niveaux la caractériser ? (QR2). Quels moyens et quelles méthodes mobiliser pour mesurer la maturité numérique des enseignants ? (QR3).

## 2.   MÉTHODE

Nous avons travaillé selon une méthode de revue herméneutique (Sackstein *et al.*, 2022), qui consiste à développer une compréhension à travers une analyse itérative et une interprétation de travaux existants sur un sujet particulier. Ce processus de revue comprend notamment la recherche, la classification, et l'évaluation critique de ces travaux (Boell et Cecez-Kecmanovic, 2014). Pour ce faire, nous avons identifié : les modèles de maturité des enseignants (1) en partant des précédentes revues systématiques de littérature portant sur les modèles d'intégration et de maturité numérique en éducation (Carvalho *et al.*, 2018 ; Franklin et Bolick, 2007 ; Harrison *et al.*, 2014 ; Kimmons *et al.*, 2020 ; Leite et Lagstedt, 2021 ; Solar *et al.*, 2013) et dans d'autres organisations (Pee et Kankanhalli, 2009 ;Teichert, 2019), (2) en suivant l'ensemble des travaux cités dans l'article ou citant l'article pour découvrir d'autres modèles jusqu'à ne plus en avoir de nouveau. Seuls les modèles applicables au contexte de l'enseignement obligatoire ont été retenus, soit 21 modèles, listés dans la figure 1 (Michel et Pierrot, 2023).

---

[2] https://pix.fr/actualites/actualite-pix-edu/

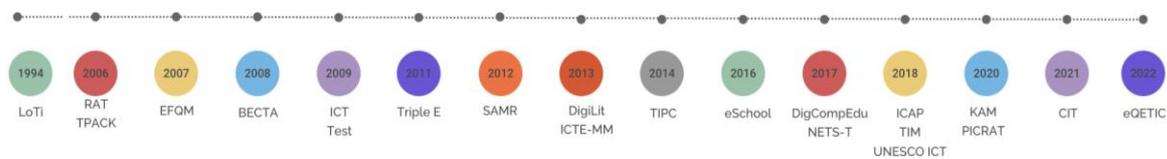

Figure 1 : Liste des 21 modèles identifiés

Nous avons ensuite comparé ces modèles en considérant plusieurs critères (Kimmons *et al.*, 2020 ; Teichert, 2019) : *le périmètre* associé à l'application du modèle (Générique G, ou Spécifique S), *la description de l'activité professionnelle* (Partielle P ou Globale G), *la place de l'apprenant* (Faible F ou Présente P), la *spécification de niveaux de maturité* (Oui O ou Non N), *l'utilité* (Accompagnement A, Descriptif De ou Diagnostic Di), *l'origine de conception du modèle* (Empirique E ou Théorique T) et *la validation* (Oui O ou Non N)). Sur cette base, nous avons choisi les 11 modèles, présentés dans la suite de l'article, qui sont : les plus génériques en termes de périmètre de description de l'activité professionnelle, les plus précis en termes de description de niveaux et qui sont basés sur des études empiriques ou qui ont été validés.

## 3. COMPARAISON DES MODÈLES SELON LEURS CARACTÉRISTIQUES DE CONCEPTION

La plupart des modèles (voir tableau 1) envisagent le contexte d'utilisation du numérique comme un élément générique, 2 modèles spécifient ce contexte. Cinq modèles ont la particularité de vouloir considérer l'ensemble de l'activité professionnelle des enseignants, en incluant les tâches en dehors de la classe (préparation, planification, etc.). L'activité d'enseignement est complétée, pour 6 modèles, par celle des apprenants. Les niveaux de maturité ne sont pas mesurés dans 4 modèles. Ces modèles ont principalement une utilité descriptive. Dans les 7 autres modèles, la maturité numérique des enseignants est considérée comme un élément de développement professionnel, d'où la présence d'outils de diagnostic, voire de guides ou feuilles de route pour favoriser le déploiement des technologies.

La modélisation de l'intégration du numérique dans l'enseignement provient, pour l'essentiel, de travaux basés sur l'observation de pratiques : 5 d'entre eux ont un ancrage théorique précisé et 7 modèles ont fait l'objet d'une validation empirique.

Tableau 1 : Synthèse des modèles en fonction de leurs principales caractéristiques

| | Périmètre | Activité professionnelle | Place de l'apprenant | Niveau de maturité | Utilité | Origine du modèle | Validation |
|---|---|---|---|---|---|---|---|
| **BECTA** (BECTA, 2008) | G | G | F | O | Di | E | N |
| **CIT** (Leit et Lagstedt, 2021) | G | G | F | N | A | E | N |
| **DigCompEdu** (Redecker, 2017) | S | G | P | O | A | T | O |
| **ICAP** (Chi et al., 2018) | G | P | P | O | A | T | O |
| **ICTE-MM** (Solar et al., 2013) | G | G | P | O | A | E | O |
| **LoTi** (Moersch, 1995 ; Stoltzfus, 2006) | G | P | F | O | Di | T | O |
| **NETS-T** (ISTE, 2017) | G | G | P | N | A | E | O |
| **PICRAT** (Kimmons et al., 2020) | S | P | P | N | De | T | N |
| **SAMR** (Puentedura, 2012) | G | P | F | O | De | E | N |
| **TIM** (Kozdras et Welsh, 2018) | G | P | P | O | Di | E | O |
| **TPACK** (Mishra et Koehler, 2006) | G | P | F | F | D | T | O |

# 4. LES MODÈLES DE MATURITÉ

Les modèles de maturité s'inscrivent dans des perspectives différentes, selon qu'ils décrivent les dynamiques d'appropriation (4.1), les dimensions mobilisées pour favoriser la maturité numérique (4.2), l'approche pédagogique mobilisée par les enseignants (4.3), la recherche d'efficacité pédagogique (4.4) et les compétences attendues (4.5) des enseignants ou qu'ils s'intéressent spécifiquement aux organisations (4.6).

## 4.1 MODÈLES BASÉS SUR LES DYNAMIQUES D'APPROPRIATION DES ENSEIGNANTS

Selon Puentedura (2012), le modèle Substitution Augmentation, Modification et Redéfinition -SAMR- encourage les éducateurs à « passer » des niveaux d'enseignement grâce à la technologie (voir figure 2a). Ce modèle fourni la structure pour considérer la manière dont l'intégration de l'outil numérique est réalisée à travers 4 étapes de transformation de la tâche d'enseignement : substitution, augmentation, modification et redéfinition de la tâche d'enseignement avec la technologie. Le modèle a été développé à partir d'observations et sans fondements théoriques, il est pourtant largement utilisé et cité dans les travaux scientifiques (Blundell *et al*., 2022).

Le modèle *Collective Integration of Technology* -CIT- (Leite et Lagstedt, 2021) considère le processus collectif de renforcement des connaissances d'un groupe (enseignants, responsables pédagogiques, décideurs) et la manière dont la culture de l'organisation peut soutenir (ou entraver) l'intégration des technologies éducatives dans les pratiques scolaires, dans une perspective interactionnelle(voir figure 2b). Le modèle identifie 4 états, plutôt qu'étapes, pour signaler que ces états, déclinés selon des aspects comportementaux, cognitifs et émotionnels ainsi que des expériences collectives, ne sont pas linéaires et peuvent être vécus simultanément par les enseignants : lorsqu'une nouvelle technologie est introduite, le collectif est d'abord dans une phase de *choc* qui précède une phase de *négociation* vis-à-vis de leurs préjugés sur l'objet (positive, s'ils jugent par exemple que la technologie peut leur faire gagner du temps, négative à l'inverse), puis dans des phases d'*autonomisation* (durant lesquelles ils construisent les premiers usages) et *exploration* (durant lesquelles ils développent de nouveaux usages).

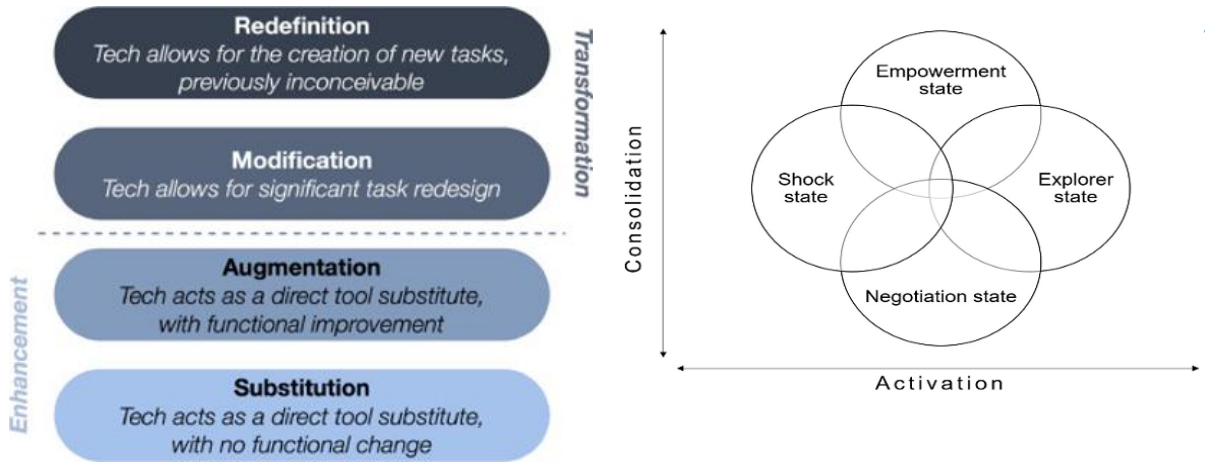

Figure 2 : (a) Modèle SAMR (Puentedura, 2012) et (b) CIT (Leite et Lagstedt, 2021)

## 4.2 Modèle basé sur les dimensions de la maturité pour les enseignants

Le modèle *Technological Pedagocial Content Knowledge* -TPACK- (Mishra et Koehler, 2006) est considéré comme l'un des modèles les plus importants décrivant les compétences des enseignants pour un enseignement réussi avec la technologie (voir figure 3). La valeur ajoutée du modèle est de ne pas considérer individuellement les connaissances technologiques (*TK*), relatives au contenu disciplinaire (*CK*) et celles pédagogiques (*PK*), mais plutôt leurs interactions matérialisées par les zones de recouvrement (*TCK*, *PCK*, *TPK* et *TPACK*). En 2019, TPACK évolue pour inclure les connaissances contextuelles (*XK*) et ainsi intégrer les contraintes organisationnelles et situationnelles liées à la mise en œuvre de la technologie en enseignement (Mishra, 2019). Le succès des efforts des enseignants ne dépend ainsi uniquement de leur connaissance *TK*, *PK*, *CK* et de leurs chevauchements, mais aussi de leur capacité à les mettre en œuvre en fonction du contexte.

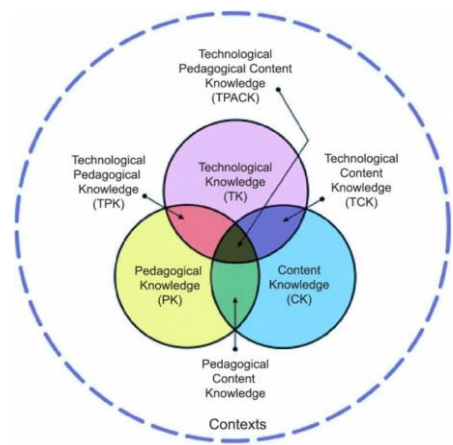

Figure 3 : Modèle TPACK avec la dimension XK (Mishra, 2019)

### 4.3 MODÈLES BASÉS SUR L'APPROCHE PÉDAGOGIQUE RETENUE PAR LES ENSEIGNANTS

Le modèle *Interactive, Constructive Active and* Passive -ICAP- (Chi *et al.*, 2018) propose une perspective différente en se concentrant sur les processus et niveaux d'engagement cognitifs des apprenants plutôt que sur le niveau de maturité ou la capacité des enseignants. Ce modèle (voir figure 4a) identifie quatre types d'activités d'apprentissage : *interactives*, *constructives*, *actives* et *passives*, en classant l'engagement cognitif du plus coûteux au moins coûteux (Antonietti *et al.*, 2023).

Les *Levels of Technology Implementation* (LoTi) (voir figure 4b) ont pour objectif d'évaluer l'efficacité de la mise en œuvre du numérique à travers 7 niveaux (du niveau 0, pour la *non-utilisation*, au niveau 6, correspondant au niveau de *raffinement*). Conceptuellement, le modèle LoTi décrit 5 dimensions (*enseignement/apprentissage avec le numérique, évaluation avec le numérique, créativité des élèves, développement professionnel* et *citoyenneté numérique*) dans une approche holistique, pour évaluer la portée complète de l'intégration des technologies. L'utilisation des outils et ressources numériques, en classe pour l'enseignement et l'apprentissage, est mesurée à l'aide d'outils validés empiriquement (Moersch, 1995 ; Stoltzfus, 2006) et vise à contribuer ensuite au développement professionnel des enseignants.

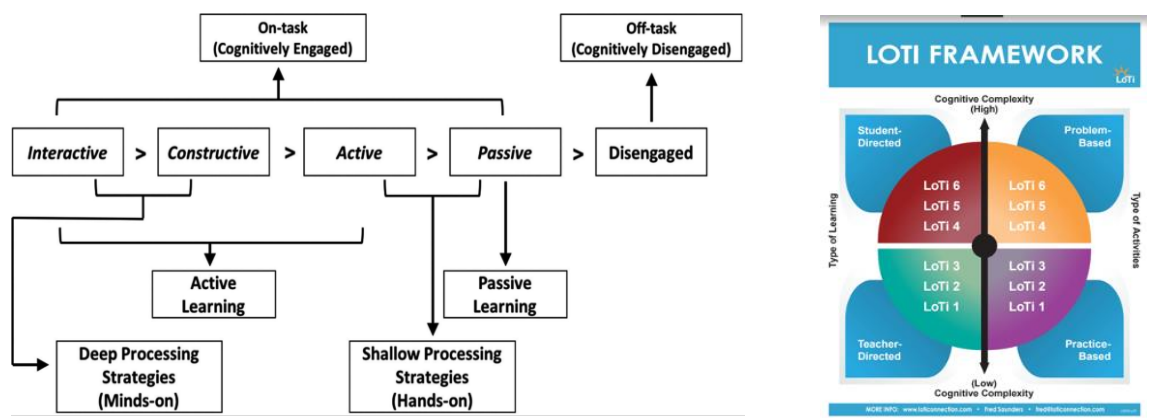

Figure 4 : (a) Modèle ICAP (Chi *et al.*, 2018) et (b) LoTi (Moersch, 1995 ; Stoltzfus, 2006)

### 4.4 MODÈLES MIXTES ARTICULANT L'EFFICACITÉ PÉDAGOGIQUE ET LES NIVEAUX DE MATURITÉ

Le modèle *Passive, Interactive, Creative, Replacement, Amplification,* Transformation (PICRAT) (Kimmons *et al.*, 2020) propose une approche centrée sur l'engagement de l'apprenant avec les outils numériques. Le modèle (voir figure 5a) se divise en deux composantes principales : l'engagement de l'apprenant avec l'outil (PIC pour *passif, interactif, créatif*) et la manière dont l'outil modifie la mise en œuvre pédagogique de l'activité (RAT pour *remplacement, amplification* ou *transformation de la pratique*), soit 9 combinaisons possibles. Pour chacune des catégories, le modèle distingue les méthodes pédagogiques, les processus d'apprentissage des élèves et les objectifs didactiques poursuivis.

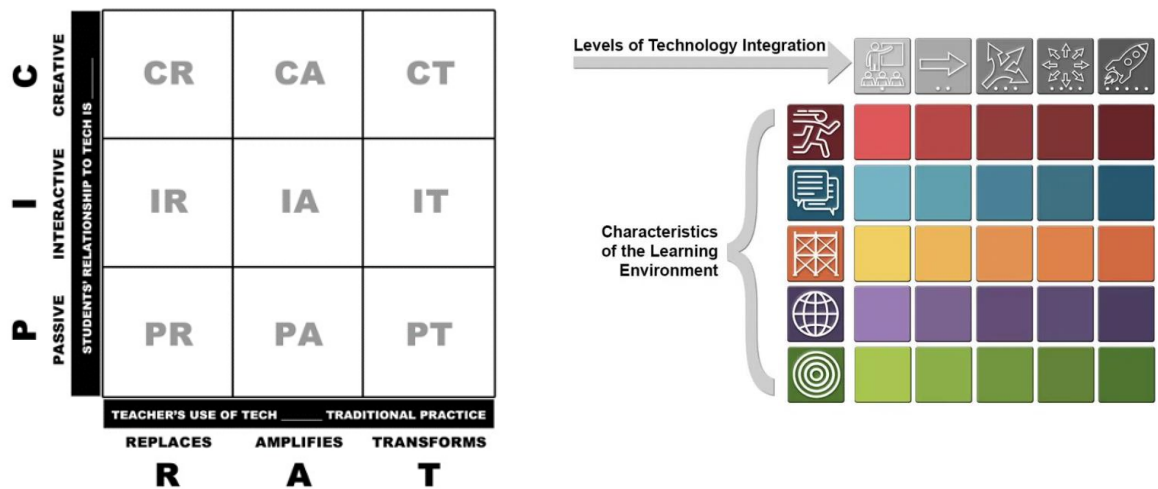

Figure 5 : (a) Modèle PICRAT (Kimmons *et al.*, 2020) et (b) TIM (Kozdras et Welsh, 2018)

La *Technology Integration Matrix* (TIM)[3] (Kozdras et Welsh, 2018) se présente sous la forme d'une matrice (voir figure 5b) destinée à évaluer le niveau d'intégration de l'outil numérique dans l'environnement d'apprentissage. Il comporte 5 niveaux d'intégration technologique (*entrée*, *adoption*, *adaptation*, *infusion* et *transformation*) et 5 caractéristiques de l'environnement d'apprentissage (*actif*, *collaboratif*, *constructif*, *authentique* et *orienté vers un objectif*). Ces caractéristiques ont été identifiées en fonction des "meilleures pratiques" identifiées par l'équipe de recherche à l'origine du modèle. Ce dernier a pour finalité d'aider l'enseignant à choisir comment utiliser les outils numériques pour atteindre les objectifs d'apprentissage.

## 4.5 MODÈLES MIXTES ARTICULANT LES COMPÉTENCES ET LES NIVEAUX DE MATURITÉ

Le *Digital Competence Framework for Educators* (DigCompEdu) (voir figure 6a), élaboré par Redecker (2017), a pour objectif de définir les compétences numériques des enseignants, pour tous les niveaux ou matières à enseigner, à l'échelle européenne. DigCompEdu considère les compétences professionnelles, pédagogiques et de l'apprenant selon *6 domaines* (eux même décomposé en 3 à 6 sous-domaines) et 6 niveaux d'utilisation du numérique dans l'éducation (du niveau A1 - *newcomer*- au niveau C2 - *pioneer*). Chaque compétence est décrite selon les activités professionnelles de l'enseignant. Sur cette base, un outil d'auto-évaluation et de diagnostic en 32 questions a été développé, à destination des enseignants et des organisations : le « SELFIE for Teachers » (Jabłonowska et Wiśniewska, 2021 cité par Tomczyk et Fedeli, 2021).

Les *National Educational Technology Standards for Teachers* (NETS-T) (voir figure 6b) se décomposent en 5 domaines *(faciliter l'apprentissage avec la technologie, concevoir des expériences d'apprentissage mobilisant la technologie, modéliser l'apprentissage, promouvoir la citoyenneté numérique, assurer son propre développement professionnel)* décrits selon 4 indicateurs (ISTE, 2017). Dans l'ensemble, ces normes sont conçues pour l'autodiagnostic et la création de programmes éducatifs permettant aux enseignants de changer les attitudes à l'égard des nouvelles technologies. Elles ont été élaborées grâce aux contributions d'acteurs divers de l'éducation (Crompton et Sykora, 2021).

---

[3] http://mytechmatrix.org et https://fcit.usf.edu/matrix/matrix

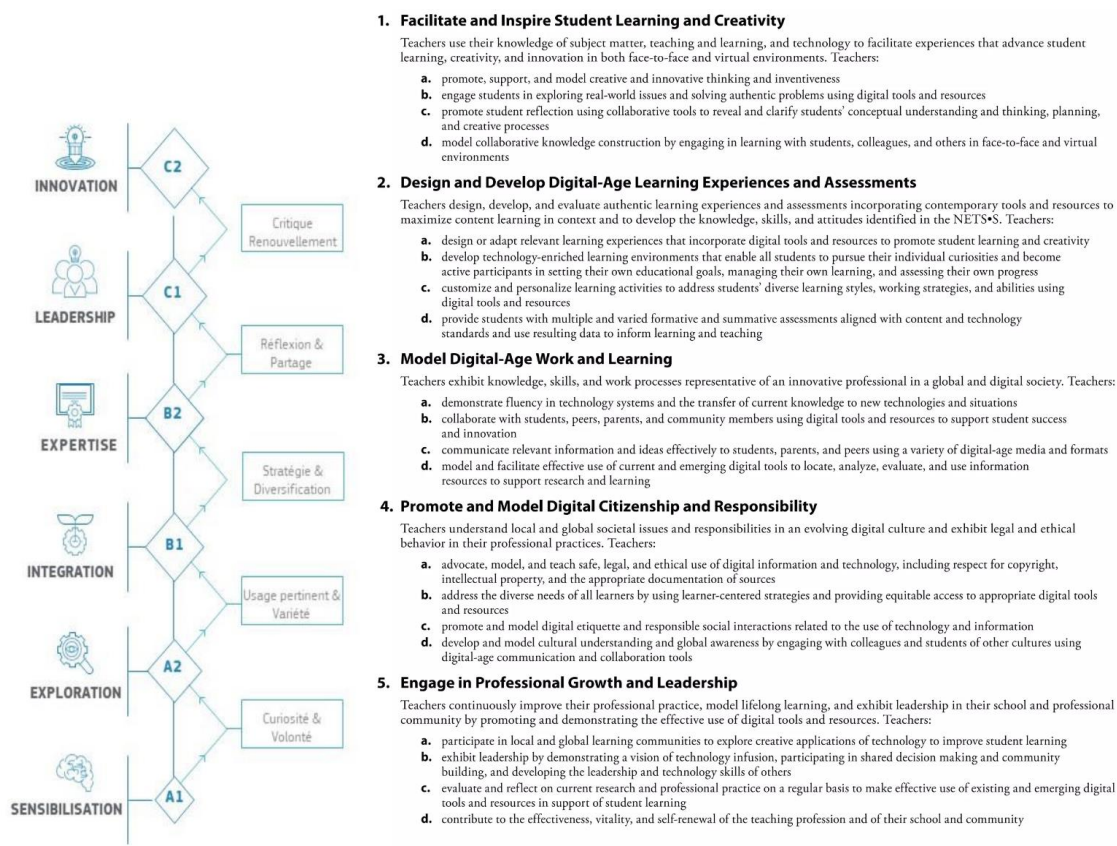

Figure 6 : (a) Modèle DigCompEdu (Redecker, 2017) et (b) NETS-T (ISTE, 2017)

## 4.6 MODÈLES DESCRIPTIFS DE LA MATURITÉ DES ORGANISATIONS

Le modèle BECTA (voir figure 7a) développé par le *British Educational Communications and Technology Agency* en 2008 est conçu pour aider les établissements d'enseignement supérieur à atteindre une maturité numérique à travers un outil d'auto-évaluation autour de 5 domaines (*leadership, contexte, ressources, soutien à l'apprentissage* et *enseignement et apprentissage*) et 5 niveaux à destination des décideurs et des enseignants (BECTA, 2008). Le modèle a été complété en 2018 (Ristić, 2018) pour décrire les contextes et cultures scolaires favorisant le développement systématisé du numérique (l'intégration) par la gestion et le soutien aux activités d'enseignement et d'apprentissage.

*ICT in School Education Maturity Model* (ICTE-MM) est une proposition (voir figure 7b) qui a pour ambition de se rapprocher de standards internationaux (Solar *et al.*, 2013) et s'inspire pour cela du modèle standardisé du CMMI[4] (*Capability Maturity Model Integration*) et des travaux de l'ISTE (2017). Le modèle distingue 3 dimensions susceptibles de soutenir les processus éducatifs (pilotage, stimulation de la culture numérique, ressources d'information et TIC). ICTE-MM propose un outil d'auto-évaluation et une feuille de route pour guider les chefs d'établissement sur la gestion du numérique.

---

[4] https://www.cmmiinstitute.com/

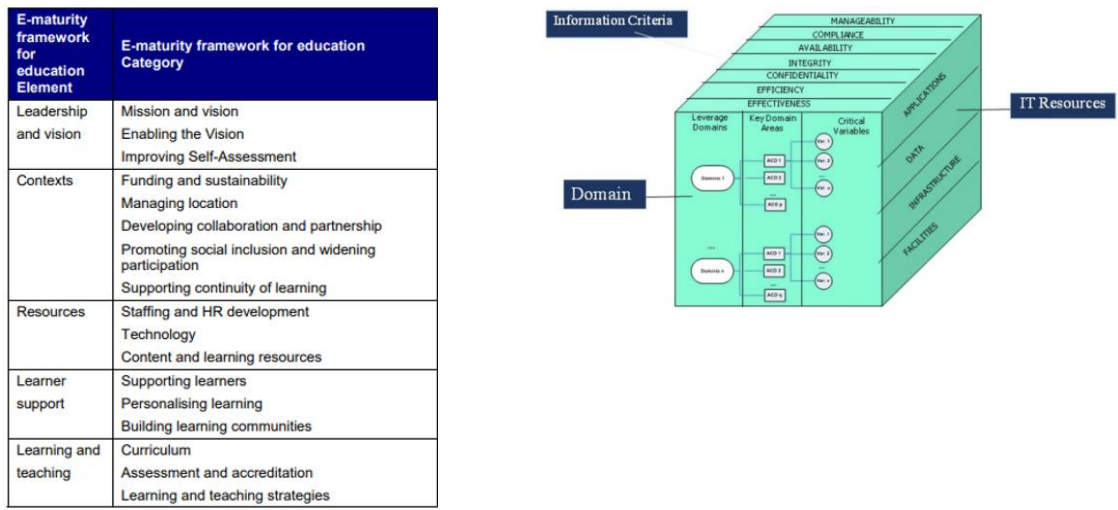

Figure 7 : (a) Modèle BECTA (BECTA, 2008) et (b) ICTE-MM (Solar *et al.*, 2013)

## 5. Vers un modèle unifié de la maturité numérique des enseignants : MUME

### 5.1 MUME : les domaines descriptifs

Les domaines de caractérisation des modèles ont été structurés en considérant les modèles d'intégration les plus généraux pour aller vers les plus spécifiques. Les modèles ont été intégrés de manière à préserver au maximum les domaines et la structure de chaque modèle. Différentes restructurations ont été opérées de manière à articuler les modèles entre eux dans une vue unifiée (voir figure 8b). Lorsqu'un domaine était déjà présent, il n'a pas été affiché dans la structuration. Ainsi l'ensemble des domaines du LoTi n'apparaissent pas explicitement dans la modélisation, mais ils sont considérés dans la mesure où ils sont déjà présents dans les autres modèles. 4 modèles structurent cette vue unifiée (voir figure 8a) : le TPACK, le ICTE-MM, le DigCompEdu et l'ICAP. Nous avons choisi de ne faire apparaitre dans le TPACK que les dimensions qui concernent l'intégration des technologies, soit TPCK et XK.

Nous avons opéré des restructurations sur l'ICTE-MM et le DigCompEdu : les élèves ont été intégrés aux domaines du DigCompEdu qui concernent l'enseignant puisque ce sont les actions de l'enseignant vers les élèves qui sont considérés et non les actions des élèves eux-mêmes. Ainsi, ces domaines sont-ils rattachés sous l'enseignant, dans la gestion des élèves. D'autres sous-domaines (du domaine 3) du DigCompEdu ont été restructurés autour : des pratiques pédagogiques de l'ICAP (pour intégrer le TIM et le PICRAT) et de la gestion des élèves (pour intégrer le sous-domaine « conseil »). La gestion de l'éducation de l'ICTE-MM a été, de la même manière, intégrée au domaine des administrateurs. Les dimensions du BECTA ont pu être ajoutées sur cette base. Les domaines du NETS-T (en jaune, à droite, dans la figure 1), ne sont pas cohérents avec les autres, car structurés par rôle plutôt que par compétence, mais ont été ajoutés pour faciliter, dans une perspective UX design, la conception de moyens ou de services d'accompagnement à la montée en maturité.

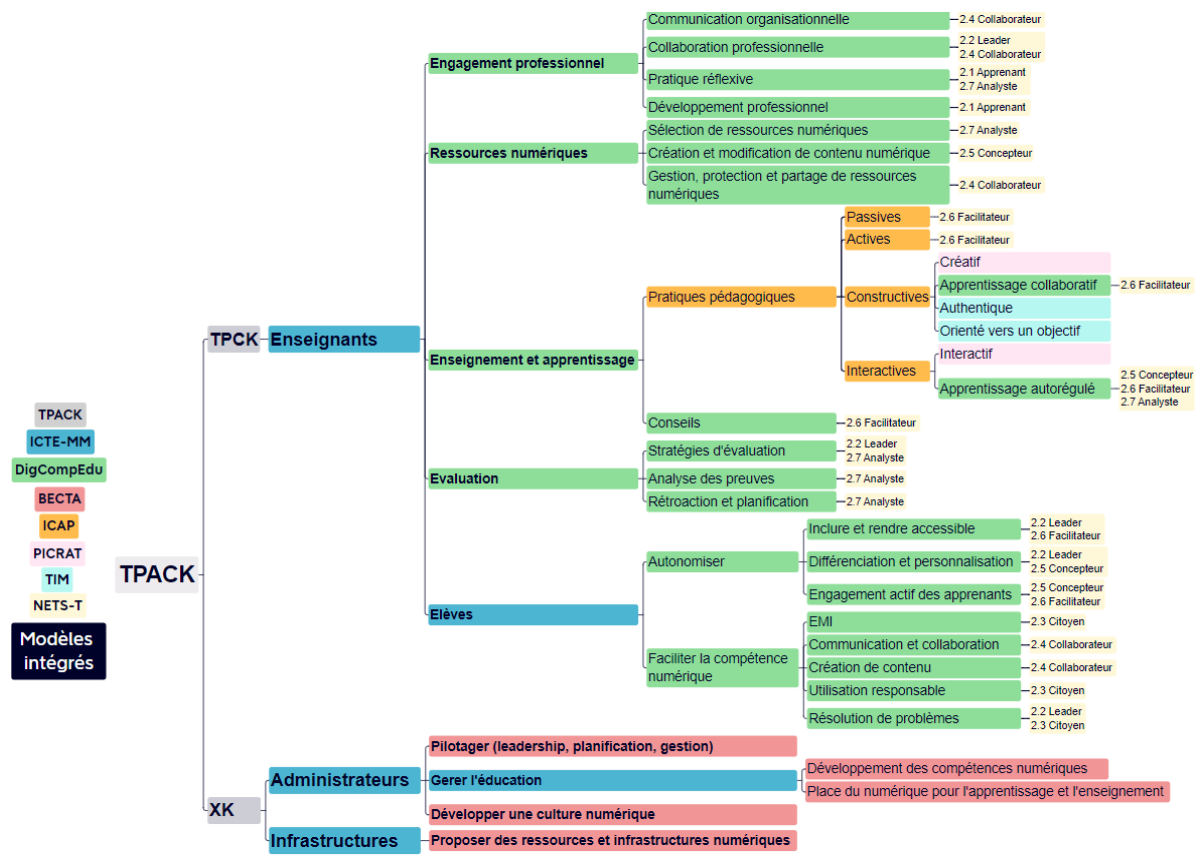

Figure 8 : (a) Modèles intégrés et (b) Critères de caractérisation sélectionnés

Le modèle unifié proposé comporte 3 domaines principaux (issus de l'ICTE-MM) : Enseignants, Administrateurs et Infrastructure. Le domaine Enseignants a été réduit à 4 sous-domaines (provenant du DigCompEdu) : engagement professionnel, ressources numériques, enseignement et apprentissage, évaluation et élèves. Les sous-domaines « enseignement et apprentissage » et « évaluation » pourraient être fusionnés, comme dans le CRCN-Edu, mais la spécificité des sous-domaines de « évaluation » qui correspond à un rôle d'analyste doit être distinguée.

## 5.2  MUME : LES NIVEAUX D'INTÉGRATION

Lorsque l'on compare les différents modèles en termes de niveau (voir tableau 2), on perçoit de nombreuses différences. Rares sont les modèles à ne pas mentionner de gradation dans les pratiques ou compétences (comme le TPACK ou les NETS-T). Les autres modèles considèrent un nombre de niveaux variant de 3 à 7. Seuls le DigCompEdu, le NETS-T et les modèles de maturité organisationnelle considèrent le rôle de leader que peut jouer l'enseignant dans la diffusion des usages et pratiques par la collaboration et le partage. Cette activité étant critique pour la diffusion des usages, nous choisissons de la conserver. De la même manière, seuls les modèles CIT, TPACK, LoTi et les modèles de maturité organisationnelle considèrent la non-utilisation. Dans la mesure où toutes les pratiques ne sont pas instrumentées et que le choix de ne pas instrumenter ses pratiques ne relèvent pas nécessairement d'un manque de compétence chez les enseignants (hors contexte COVID), mais plutôt d'un choix pédagogique, nous conservons cette catégorie et intégrons dans la population un groupe de non-utilisateurs.

Tableau 2 : Synthèse des modèles en fonction des niveaux de maturité

| Modèles | Description des niveaux | | | | | | Nb niveaux |
|---|---|---|---|---|---|---|---|
| **DOI** | Innovateur | Adoptant précoce | Majorité précoce | Majorité acquise | Retardataire | | 5 |
| **ICTE-MM** | 5 Optimisé. | 4 Géré | 3 Défini | 2 Développement | 1 Initial | | 5 |
| **BECTA** | 5 Maturité | 4 Avancé | 3 Compétent | 2 Autonomisation | 1 Pas de maturité | | 5 |
| **DigComp Edu** | Pionnier (C2) | Leader (C1) | Expert (B2) | Intégrateur (B1) | Explorateur (A2) | Nouvel arrivant (A1) | 6 |
| **LoTi** | Perfectionnement | Expansion | Intégration | Infusion | Exploration | Sensibilisation | Non-utilisation — 7 |
| **ICAP** | | | Interactif | Collaboratif | Actif | Passif | 4 |
| **PICRAT** | Transformation | | Amplification | | Remplacement | | 3 |
| **TIM** | Transformation | | Infusion | Adaptation | Adoption | Entrée | 5 |
| **SAMR** | Redéfinition | Modification | | Augmentation | Substitution | | 4 |
| **CIT Model** | Explorateur | Autonomisation | | | Négociation | Choc | 4 |
| **TPACK** | TPCK | | | | | TK, PK, CK | 1 |
| **NETS-T** | | | | | | | Pas de niveau |
| **Synthèse** | Transformation | Développement | Intégration | Amélioration | Substitution | Non-utilisation | 6 |
| | Pionnier | Leader | Expert | Explorateur | Nouvel arrivant | Non-utilisateur | 6 |

Les modèles ont aussi différents points communs. À part le NET-S, tous intègrent une gradation de la maturité allant d'un niveau « entrée », qui correspond aux usages les plus simples, à un niveau « transformation », correspondant à la création d'innovation d'usage avec la technologie. Dans la plupart des modèles (SAMR, CIT, ICAP, LoTi, TIM, DigCompEdu, BECTA, ICTE-MM) cette gradation considère l'expertise en termes de compétences techno-pédagogiques avec un cœur à 4 niveaux, globalement aligné sur les définitions de l'ICAP (passif, actif, collaboratif, interactif) et un niveau 5 qui correspond à la capacité à innover vers de nouvelles formes techno-pédagogiques.

Si on ajoute un niveau de « non-utilisation » au modèle de diffusion de l'innovation (ou DOI) de Rogers (2003), on peut constater que les niveaux 6, 5, 3, 2, 1 (« Innovateur », « Adoptant précoce », « Majorité acquise », « Retardataire » et « Non-utilisateur ») sont cohérents pour l'ensemble des modèles. Les niveaux 1 et 2 correspondent respectivement à une non-maturité et une entrée dans le processus d'intégration des technologies principalement par des pratiques de conception simple et de transmission de supports de formation. Le niveau 3 (majorité active) correspond à une phase d'exploration des possibilités et se concrétise au niveau 4 (et majorité précoce) par des stratégies pédagogiques actives. Le niveau 5 est plutôt caractérisé par des pratiques de leadership et de partage vers les autres membres de la communauté, ainsi que de gestion et d'analyse. Le niveau 6 est caractérisé par des capacités d'innovation et de maîtrise complète de l'intégration des technologies. Le niveau 4 est moins cohérent. Il est souvent distingué dans les modèles spécifiques pour l'éducation en 2 niveaux : (expert, intégrateur) pour le DigCompEdu, (Infusion, Intégration) pour le LoTi et (Infusion, Adaptation) pour le TIM. Cette distinction ne nous semble utile que pour identifier les pratiques interactives et collaboratives. En effet, les pratiques interactives sont actuellement peu développées et pourraient correspondre au niveau « adoptant précoce », mais ne correspondent pas à la capacité de diffusion et leadership de cette catégorie. Nous choisissons donc dans un premier temps de les intégrer au niveau 4 et vérifierons la cohérence de ce choix par une étude empirique.

Nous recommandons d'utiliser un modèle en 6 niveaux qui considère soit les processus caractéristiques (Transformation, Développement, Intégration, Amélioration, Substitution, Non-utilisation), soit les pratiques des acteurs (Pionnier, Leader, Expert, Explorateur, Nouvel arrivant, Non-utilisateur). La courbe correspondante est représentée dans la figure suivante (voir figure 9), à titre indicatif. Des études empiriques doivent être menées pour en définir la forme précise. La classification de Rogers (2003) y a aussi été présentée à titre de comparaison.

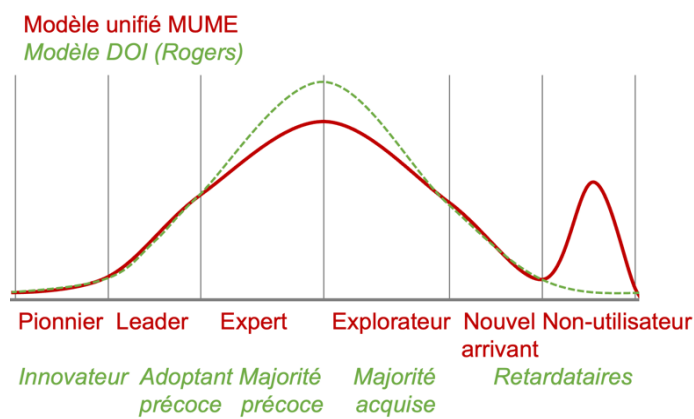

Figure 9 : Comparaison des courbes de diffusion des technologies pour l'éducation pour le
DOI de Rogers (2003) et le modèle unifié MUME.

## 6. VERS UN DIAGNOSTIC UNIFIÉ DE LA MATURITÉ NUMÉRIQUE

Nos travaux sur l'analyse de la maturité numérique se sont poursuivis par l'étude des moyens de mesure de cette maturité. En effet, la plupart des modèles intègrent un outil de diagnostic permettant d'évaluer la maturité numérique des individus. Notre objectif était double : comparer les différents moyens pour vérifier leur cohérence et construire un outil de ce type pour faire des diagnostics adaptés pour le modèle MUME.

### 6.1 LES OUTILS DE DIAGNOSTICS SOURCES

La revue de littérature a permis d'identifier et sélectionner trois outils évaluatifs utilisés par certains modèles constitutifs de MUME : à savoir l'ICAP-TS (Antonietti *et al.*, 2023), le TPACK.x (Dorsaz, 2022 ; Schmid *et al.*, 2020), et le SELFIE (Costa *et al.*, 2021). Nous avons aussi utilisé le questionnaire TNE 95-02 développé dans une précédente recherche (Michel et Pierrot, 2022).

Le TPACK.x correspond à un questionnaire d'auto-évaluation comprenant 36 items en fonction des 8 domaines du TPACK, dont celui correspondant au contexte. Le TPACK.x est une traduction francophone validée (Dorsaz, 2022) du questionnaire existant TPACK.xs (Schmid *et al.*, 2020) et qui utilise une échelle d'accord en 5 niveaux (de 1 « fortement en désaccord à 5 « tout à fait d'accord »). Nous n'avons retenu que les items correspondant au domaine de connaissances XK (le contexte) répartis en 3 sous-domaines : $XK_{ma}$ pour les connaissances contextuelles au niveau macro, $XK_{me}$ pour celles au niveau méso et $XK_{mi}$ pour celles au niveau micro. Le niveau macro vise à appréhender les principales dimensions pertinentes dans l'intégration du numérique et concerne les actions politiques, culturelles, économiques et stratégiques relatives au déploiement des technologies. Le niveau micro se situe à l'échelle de la classe où se déploie l'activité d'enseignement, il tient compte des caractéristiques des apprenants et des enseignants. Entre ces deux niveaux, le niveau méso décrit l'environnement qui encadre la classe (prise en compte des pratiques pédagogiques au sein de l'établissement ou des pratiques des familles par exemple).

Le questionnaire d'auto-évaluation ICAP-TS (Antonietti *et al.*, 2023) est décomposé en 4 sous-échelles en fonction de la taxonomie ICAP pour évaluer des activités d'apprentissage de type passif, actif, constructif ou interactif, soutenues par le numérique. La sous-échelle « passif » concerne la mise en œuvre d'activités de type transmissives avec le numérique. La sous-échelle « actif » vise les activités qui s'appuie sur l'utilisation par les élèves du numérique pour appliquer des connaissances. La sous-échelle « constructif » décrit des activités d'apprentissage avec le numérique où les élèves construisent de nouvelles connaissances. La sous-échelle « interactif » concerne la mise en œuvre d'activités avec le numérique de type collaboratives. En complément des 4 sous-échelles, le questionnaire ICAP-TS inclut des questions relatives à la fréquence d'utilisation de 12 outils numériques (logiciel de traitement de texte ou service de test en ligne par exemple). Le questionnaire s'appuie sur une échelle de fréquence en 5 points (0 - Presque jamais, 4 - Presque que tous les jours). Le questionnaire est proposé en anglais, il a fait l'objet d'une traduction.

L'outil d'auto-évaluation SELFIE (Costa *et al.*, 2021) est adapté au DigCompEdu et inclut trois questionnaires, dont un pour les enseignants. Ce questionnaire, multilingue, est composé de plusieurs éléments pour collecter les attitudes et croyances des répondants, leurs pratiques et compétences numériques ainsi que des informations sur les ressources mises à leur disposition et leur contexte d'enseignement. Pour l'essentiel des questions, le questionnaire repose sur une échelle de fréquence en 5 points (1 - pas du tout d'accord à 5 - tout à fait d'accord). Les items relatifs à l'auto-évaluation de la maîtrise des compétences numériques sont calqués sur le référentiel de compétences DigCompEdu, avec une échelle en 7 points (A0 - je ne sais pas comment faire à C2 - je conçois de nouvelles manières de le faire).

Enfin, le questionnaire TNE 95-02 (Michel et Pierrot, 2022) est organisé en 4 parties pour identifier le contexte professionnel d'enseignement des répondants, leurs pratiques numériques avant, pendant et après la période de crise sanitaire, leur expérience d'enseignement durant le confinement et leurs caractéristiques personnelles. Ce questionnaire présente l'intérêt d'être adapté au contexte d'enseignement français, dans la mesure où il a été conçu pour être diffusé auprès d'enseignants des 1er et 2nd degrés impliqués dans le programme Territoire numérique éducatif.

## 6.2 LE DÉVELOPPEMENT D'UN NOUVEAU QUESTIONNAIRE

Le développement du nouveau questionnaire pour mesurer les niveaux de maturité a consisté à identifier les groupes de questions mobilisables dans les outils d'évaluation décrits précédemment, et cela de façon complémentaire. La fusion de l'ensemble des outils a permis de produire une première version du questionnaire incluant 23 questions regroupées en 3 parties :

- caractéristiques personnelles (14 questions à choix unique ou multiple, provenant du questionnaire TNE 95-02) ;
- utilisation du numérique (8 questions de type échelle de fréquence ou d'accord, provenant du TPACK.x et ICAP-TS) ;
- compétences numériques (1 question de type échelle en 7 points provenant du SELFIE).

À ce stade du développement du questionnaire, les questions du TPACK.x, du SELFIE et de TNE 95-02 étant déjà disponibles en français, seuls les items de l'échelle ICAP-Ts ont fait l'objet d'une première traduction et adaptation culturelle. Les items ont d'abord été traduits de l'anglais au français puis adaptés de sorte à être applicables au contexte français.

Le questionnaire élaboré a ensuite fait l'objet d'un prétest qualitatif. La compréhensibilité et la pertinence des questions ont été discutées, entre les mois de mars et d'avril 2023, avec 6 enseignants-chercheurs disposant tous d'une expérience solide dans l'accompagnement au déploiement de dispositifs numériques dans les $1^{er}$ et $2^{nd}$ degrés. Ce prétest a permis de mettre en évidence la complexité associée à la compréhension de l'échelle TPACK.x. Plus spécifiquement, les questions relatives à la dimension macro ($XK_{ma}$) ont été perçues par l'ensemble des experts comme inadaptées au contexte socioculturel français (par exemple les questions du type « *Je suis au courant de la culture régionale et nationale en matière d'utilisation des technologies pour l'enseignement* »). Le deuxième retour a porté sur la variété et le manque d'homogénéité des échelles mobilisées, celle du TPACKx et ICAP-TS étant en 5 points avec un premier niveau à 0 ou à 1, celles du SELFIE en 7 points. Le troisième retour porte sur la traduction de certains items de l'échelle ICAP-TS, jugée inadaptée. C'est le cas par exemple du libellé « *pour qu'ils construisent individuellement de nouvelles connaissances* », jugé trop complexe. Enfin, dans l'ensemble, les experts ont trouvé certaines questions des parties A et B redondantes et le questionnaire trop long pour être acceptable.

Le questionnaire a été retravaillé pour simplifier la formulation de certaines questions sur la base des retours des experts. Nous avons néanmoins conservé certaines redondances dans les questions de manière à pouvoir comparer et évaluer la cohérence des résultats selon les quatre questionnaires sources.

La nouvelle version du questionnaire inclut toujours 3 parties :

- caractéristiques personnelles (7 questions à choix unique ou multiple, provenant du questionnaire TNE 95-02) ;
- utilisation du numérique (7 questions de type échelle de fréquence ou d'accord en 7 points, provenant du TPACK.x et ICAP-TS) ;
- compétences numériques (1 question de type échelle en 7 points provenant du SELFIE).

## 6.3  ÉVALUATION PRÉLIMINAIRE DU QUESTIONNAIRE

Dans l'objectif de vérifier sa validité, le nouveau questionnaire a été diffusé au mois de juin 2023 dans le cadre du projet CoAI-Datastim auprès d'enseignants de l'académie de Paris des 1er et 2nd degrés via l'ENT PCN[5] opéré par l'entreprise EDIFICE[6]. L'intérêt de cette collecte de données est qu'en complément du questionnaire, nous disposons, par EDIFICE, des traces d'activité sur les ENT PCN. Nous pourrons ainsi comparer et mettre en perspective les usages déclarés concernant l'ENT dans le questionnaire avec les usages effectifs de l'ENT. L'analyse des données est en cours. Nous proposons ici quelques résultats préliminaires.

L'échantillon analysé comprend 143 participants, répartis entre 86 enseignants du 1er degré, 18 participants du 2nd degré, et 39 participants sans réponse. Les femmes constituent la majorité (101), comparées aux 22 enseignants hommes et aux 20 participants sans réponse. Dans l'échantillon, les enseignants de 41 à 50 ans sont les plus nombreux (52), suivis par ceux de plus de 51 ans (43). Les tranches d'âge inférieures à 30 ans et de 31 à 40 ans comptent respectivement 6 et 23 enseignants. Treize participants n'ont pas complété cette question. La majorité des enseignants (88) ont plus de 10 ans d'expérience, tandis que 23 en ont entre 3 et 10 ans. Cinq enseignants ont moins de 3 ans d'expérience, et 8 enseignants ont une expérience qualifiée comme « Autre ». Les données d'expérience ne sont pas disponibles pour 19 enseignants.

La validité du questionnaire a été évaluée en s'appuyant sur la mesure de la fiabilité des questions, à travers le coefficient α de Cronbach. La valeur de ce coefficient s'établit généralement entre 0 et 1 pour des sous-échelles ou groupes de questions et est considérée comme *acceptable* à partir de 0,70. L'ensemble des questions de notre questionnaire obtient un score supérieur à cette valeur seuil (voir tableau 3), mais il convient de préciser que notre échantillon est ici restreint : il est préconisé de collecter des données auprès d'un échantillon de 200 à 300 participants pour appréhender l'hétérogénéité des réponses.

Tableau 3 : α de Cronbach des groupes de questions relatives aux différents thèmes du questionnaire

| Groupes de variables | α de Cronbach |
|---|---|
| Utilisation du numérique en classe | 0,959 |
| Utilisation du numérique par les élèves (ICAP-TS) | 0,945 |
| Utilisation du numérique par les enseignants (ICAP-TS) | 0,922 |
| Utilisation du numérique et prise en compte du contexte de la classe XKmi (TPACK.x) | 0,906 |
| Utilisation du numérique et prise en compte du contexte de l'environnement XKme (TPACK.x) | 0,709 |
| Compétences numériques (SELFIE) | 0,967 |

L'analyse préliminaire des résultats révèle en outre des indices sur les dynamiques d'intégration du numérique dans les pratiques pédagogiques des enseignants et plus globalement leur montée en maturité. Comme nous l'avions fait précédemment (Michel *et al.*, 2021 ; Michel et Pierrot, 2022), nous avons identifié des classes de maturité des répondants selon leur auto-évaluation des compétences numériques, à l'aide d'une classification K-Means en 7 niveaux. De manière à comparer les mesures de chaque modèle, nous avons opéré cette classification sur chaque sous-groupe de questions liées aux questionnaires sources, par exemple CDig pour les questions liées au SELFIE ou CEnclasse pour les questions de l'ICAP-TS. L'analyse comparée des différentes classifications est en cours, nous la présenterons ultérieurement. À titre préliminaire nous présentons dans le paragraphe suivant les résultats de la classification réalisée sur la base des questions extraites du SELFIE.

---

[5] https://www.parisclassenumerique.fr/
[6] https://edifice.io

Les 7 classes, CDig1 à CDig7, listées dans la première ligne de le tableau 4, permettent d'identifier des dynamiques selon les usages et compétences décrits dans la première colonne du tableau. Ces usages et compétences vont de 0 (je ne sais pas comment faire) à 4 (je le fais régulièrement et je conçois de nouveaux moyens pour le faire), ils sont classés selon un niveau d'usage global (dernière colonne). La première classe d'enseignants (CDig1, 37 enseignants) correspond à ceux qui déclarent le moins savoir utiliser le numérique, qui sert essentiellement des objectifs de communication (rattachés au domaine 1 du DigCompEdu). Les classes CDig2 à 5 (respectivement, 1, 17, 8 et 2 enseignants) ont une utilisation plus diversifiée et régulière du numérique que la CDig1. Leurs usages et compétences s'inscrivent autour des domaines 2 (ressources numériques), 6 (développement des compétences numériques des apprenants) et 3 (enseignement et apprentissage). La progression d'une classe à l'autre est guidée par les items réflexifs et d'autoformation. La classe CDig6, 2 enseignants, et la classe CDig7, 4, correspondent aux enseignants dont les usages sont les plus développés, y compris dans les domaines du DigCompEdu d'évaluation et d'autonomisation des apprenants.

Tableau 4 : Classes d'enseignants selon leur maîtrise des compétences numériques

| Domaine | Classe DigComp Ens/Usages | CDig1 | CDig2 | CDig3 | CDig4 | CDig5 | CDig6 | CDig7 | Niveau d'usage |
|---|---|---|---|---|---|---|---|---|---|
| D4 | [Utiliser les outils numériques pour évaluer l'apprentissage de manière formative et sommative] | 0,01 | 0,00 | 0,24 | 0,58 | 0,50 | 0,70 | 2,05 | 0,58 |
| D5 | [Utiliser les outils numériques pour s'adapter aux différents niveaux et rythmes d'apprentissage des élèves] | 0,01 | 0,00 | 0,73 | 0,10 | 0,10 | 0,90 | 2,75 | 0,65 |
| D2 | [Organiser les contenus numériques pour qu'ils soient faciles à utiliser pour les élèves, les parents et les enseignants] | 0,05 | 0,00 | 0,91 | 0,68 | 0,10 | 0,50 | 2,50 | 0,68 |
| D3 | [Utiliser des technologies innovantes (Robotique, IA, RV) pour explorer de nouveaux contenus et expériences d'apprentissage] | 0,01 | 0,00 | 0,12 | 0,53 | 0,10 | 2,40 | 2,10 | 0,75 |
| D3 | [Utiliser les outils numériques pour aider les élèves à résoudre des problèmes et mieux comprendre] | 0,08 | 0,00 | 0,35 | 0,10 | 0,10 | 3,00 | 2,75 | 0,91 |
| D3 | [Utiliser les outils numériques pour aider les élèves à apprendre de manière autonome] | 0,05 | 0,00 | 0,46 | 0,20 | 0,10 | 3,00 | 2,75 | 0,94 |
| D4 | [Utiliser les outils numériques pour suivre les progrès individuels des élèves] | 0,01 | 0,00 | 0,29 | 0,63 | 0,50 | 2,40 | 2,75 | 0,94 |
| D5 | [Tenir compte des facteurs affectifs des élèves (motivation, fatigue, etc.) pour garantir un environnement d'apprentissage favorable] | 0,00 | 0,00 | 0,72 | 1,13 | 1,00 | 1,50 | 2,25 | 0,94 |
| D3 | [Utiliser les outils numériques pour que les élèves travaillent ensemble et communiquent] | 0,06 | 0,00 | 0,35 | 0,08 | 0,10 | 4,00 | 2,50 | 1,01 |
| D6 | [Donner aux élèves les moyens d'utiliser les outils numériques en toute sécurité (gestion des risques liés à leur bien-être physique, psychologique et social)] | 0,01 | 0,00 | 0,66 | 0,35 | 0,10 | 3,50 | 2,50 | 1,04 |
| D5 | [Tenir compte des conditions d'accès aux outils numériques (disponibilité matérielle, compétences, etc.) pour garantir un environnement d'apprentissage favorable] | 0,00 | 0,00 | 0,60 | 0,50 | 1,00 | 2,50 | 1,75 | 1,05 |
| D3 | [Utiliser les outils numériques pour que les élèves produisent des contenus et s'expriment] | 0,06 | 0,00 | 0,35 | 0,20 | 0,30 | 3,50 | 3,00 | 1,06 |
| D6 | [Donner aux élèves les moyens d'utiliser les outils numériques de manière responsable et éthique (gérer leur identité numérique, empreinte numérique, réputation numérique, etc.)] | 0,01 | 0,00 | 0,68 | 0,13 | 0,10 | 4,00 | 2,75 | 1,09 |
| D5 | [Tenir compte des caractéristiques socio-démographiques des élèves (âge, milieu socio-économique, culture, langue maternelle, etc.) pour garantir un environnement d'apprentissage favorable] | 0,02 | 0,00 | 0,71 | 0,88 | 1,00 | 2,50 | 2,75 | 1,12 |
| D2 | [Utiliser les outils numériques pour créer ou modifier des ressources numériques pour mes cours] | 0,14 | 0,00 | 1,02 | 0,20 | 3,50 | 0,90 | 2,75 | 1,22 |
| D2 | [Utiliser les outils numériques pour trouver et sélectionner des ressources numériques pour mes cours] | 0,24 | 0,00 | 1,00 | 0,30 | 3,50 | 1,90 | 2,25 | 1,31 |
| D3 | [Utiliser les outils numériques pour aider les élèves à apprendre plus efficacement] | 0,29 | 0,00 | 0,89 | 0,73 | 0,80 | 4,00 | 2,50 | 1,32 |
| D1 | [Utiliser les outils numériques pour aider les élèves et les enseignants à réfléchir sur leur travail] | 0,01 | 3,00 | 0,44 | 0,15 | 0,10 | 4,00 | 2,50 | 1,46 |
| D2 | [Respecter les lois sur la propriété intellectuelle lors de l'utilisation des outils numériques] | 0,05 | 0,00 | 1,62 | 1,48 | 2,50 | 3,50 | 2,05 | 1,60 |
| D1 | [Utiliser les outils numériques pour travailler avec mes collègues et d'autres personnes impliquées dans l'éducation] | 0,39 | 3,00 | 1,93 | 1,39 | 2,25 | 1,90 | 0,70 | 1,80 |
| D1 | [Utiliser les outils numériques pour apprendre de nouvelles choses pour mon travail] | 0,16 | 0,00 | 1,27 | 0,90 | 3,50 | 4,00 | 3,00 | 1,83 |
| D2 | [Protéger les informations personnelles des élèves, des parents et des enseignants en ligne] | 0,10 | 3,00 | 0,96 | 2,25 | 2,50 | 3,00 | 2,75 | 2,08 |
| D1 | [Réfléchir à la manière dont j'utilise les outils numériques dans mon travail] | 0,09 | 4,00 | 0,80 | 2,10 | 2,50 | 3,00 | 2,50 | 2,14 |
| D1 | [Utiliser les outils numériques dans mon école pour communiquer avec les élèves, les collègues et les parents] | 0,66 | 3,00 | 1,56 | 2,48 | 2,40 | 1,90 | 3,00 | 2,14 |
| | Nb Ens | 37,00 | 1,00 | 17,00 | 8,00 | 2,00 | 2,00 | 4,00 | |
| | Variance intra-classe | 2,45 | 0,00 | 10,02 | 23,57 | 17,74 | 28,18 | 16,41 | |
| | Moyenne | 0,10 | 0,67 | 0,76 | 0,79 | 1,18 | 2,55 | 2,60 | |

# 7. Conclusion et perspectives

Cet article propose une revue de littérature sur la maturité numérique des enseignants. Nous avons identifié 11 modèles : 9 modèles spécifiques aux pratiques professionnelles des enseignants (sections 2.1 à 2.5) et 2 modèles portant sur leur contexte professionnel (section 2.6). Sur la base d'une analyse comparée de ces modèles, nous proposons un modèle unifié MUME, considérant les aspects individuels liés à l'enseignant, et les aspects organisationnels et contextuels. Ce choix permet de l'utiliser pour des travaux globaux sur l'intégration du numérique en éducation ou pour d'autres considérants uniquement l'enseignant. En outre, ce modèle présente l'avantage de couvrir l'ensemble de l'activité professionnelle de l'enseignant, plutôt qu'uniquement ses tâches d'enseignement. Le modèle unifié est composé de 6 niveaux cohérents globalement avec ceux du DigCompEdu, du DOI de Rogers et du ICTE-MM. Il a, en outre, la particularité d'intégrer un niveau de maturité 0 (niveau 1), correspondant à une non-utilisation que nous considérons comme un choix de l'enseignant plutôt qu'un frein, et de fusionner les niveaux B1 et B2 du DigCompEdu. Ce choix se justifie par le fait de proposer un outil mobilisable, à terme, pour du diagnostic et de l'accompagnement à l'intégration du numérique.

Notre modèle unifié de maturité constitue une première contribution à l'observation et l'analyse des niveaux de maturité numérique des enseignants. Les questionnaires identifiés grâce à la revue de littérature présentent l'intérêt majeur d'offrir une vue des niveaux de maturité des enseignants. Cependant, ces moyens sont tous basés sur des données auto-rapportées et correspondent à un diagnostic à un temps donné. De manière à pouvoir comprendre la progression des usages (Michel et Pierrot, 2022), une approche mixte d'observation et d'analyse des niveaux de maturité numérique des enseignants parait pertinente. De cette manière, ce travail se poursuit actuellement par le développement d'un outil de diagnostic de la maturité basé sur un questionnaire. À ce stade, il nous reste à poursuivre l'analyse des données pour en vérifier la validité. Plus globalement, cette mise en application du modèle MUME, orientée vers la collecte de données déclaratives, sera combinée à d'autres types de données, collectées automatiquement (logs d'usage des outils numériques) ou non (enregistrements vidéo d'activités d'enseignement). Notre objectif est, à terme, de nous appuyer sur les potentialités offertes par les travaux des *Learning* et *Teaching Analytics* pour proposer un tableau de bord de la maturité numérique à destination des enseignants.

# 8. Remerciements

# 9. Bibliographie